\documentstyle[12pt,epsf]{article}

\begin{document}
\baselineskip 0.7cm

\newcommand{\beq}{ \begin{eqnarray} }
\newcommand{\eeq}{ \end{eqnarray} }
\newcommand{\beqstar}{ \begin{eqnarray*} }
\newcommand{\eeqstar}{ \end{eqnarray*} }
\newcommand{\gsim}{ \mathop{}_{\textstyle \sim}^{\textstyle >} }
\newcommand{\lsim}{ \mathop{}_{\textstyle \sim}^{\textstyle <} }
\newcommand{\vev}[1]{ \left\langle {#1} \right\rangle }
\newcommand{\lsp}{ \left ( }
\newcommand{\rsp}{ \right ) }
\newcommand{\lmp}{ \left \{ }
\newcommand{\rmp}{ \right \} }
\newcommand{\llp}{ \left [ }
\newcommand{\rlp}{ \right ] }
\newcommand{\labs}{ \left | }
\newcommand{\rabs}{ \right | }
\newcommand{\EV}{ {\rm eV} }
\newcommand{\KEV}{ {\rm keV} }
\newcommand{\MEV}{ {\rm MeV} }
\newcommand{\GEV}{ {\rm GeV} }
\newcommand{\TEV}{ {\rm TeV} }
\newcommand{\mgut}{M_{\rm GUT}}
\newcommand{\mint}{M_{I}}
\newcommand{\mll}{m_{\tilde{l}L}^{2}}
\newcommand{\mdr}{m_{\tilde{d}R}^{2}}
\newcommand{\mllXX}[1]{m_{\tilde{l}L , {#1}}^{2}}
\newcommand{\mdrXX}[1]{m_{\tilde{d}R , {#1}}^{2}}
\newcommand{\mgy}{m_{G1}}
\newcommand{\mgl}{m_{G2}}
\newcommand{\mgc}{m_{G3}}
\newcommand{\nuR}{\nu_{R}}
\newcommand{\slL}{\tilde{l}_{L}}
\newcommand{\slLi}{\tilde{l}_{Li}}
\newcommand{\sdR}{\tilde{d}_{R}}
\newcommand{\sdRi}{\tilde{d}_{Ri}}
\newcommand{\djn}{N=$\hbox{2\kern-.45em /}$ }

\renewcommand{\thefootnote}{\fnsymbol{footnote}}
\setcounter{footnote}{1}

\begin{titlepage}

\begin{flushright}
NSF-ITP-95-124\\
TIT-HEP-303\\
UT-726\\
October, 1995
\end{flushright}

\vskip 0.35cm
\begin{center}
{\large \bf
An \djn SUSY Gauge Model for Dynamical Breaking \\
of the Grand Unified SU(5) Symmetry}
\vskip 1.2cm
J.~Hisano$^{\it a,b}$ and T.Yanagida$^{\it a,c}$
\vskip 0.4cm
{\it
 $a)$~Institute for Theoretical Physics, University of California,\\
 ~~~Santa Barbara, CA 93106, U.S.A.\\
 $b)$~Department of Physics, Tokyo Institute of Technology,\\
 ~~~Oh-okayama 2-12-1, Megro-ku, 152, Japan\\
 $c)$~Department of Physics,  University of Tokyo,\\
 ~~~Hongo 7-3-1, Bunkyo-ku, 113, Japan}

\vskip 1.5cm

\abstract{
We construct an extension of the recently proposed dynamical model for the
breaking of SU(5)$_{\rm GUT}$ gauge symmetry, in which a pair of massless
chiral supermultiplets for Higgs doublets are naturally obtained. We point
out that a model at a specific point in the parameter space of superpotential
is regarded as a low-energy effective theory of an \djn supersymmetric gauge
model for
the strongly interacting hypercolor sector.
}
\end{center}
\end{titlepage}

\renewcommand{\thefootnote}{\arabic{footnote}}
\setcounter{footnote}{0}

%
%
%
%

Recently, a dynamical breaking scenario of the grand unified (GUT) gauge
symmetry has been proposed based on a supersymmetric (SUSY) hypercolor
SU(3)$_{\rm H}$ gauge theory, where six flavors of quarks $Q$ and
antiquarks $\overline{Q}$ interact strongly at the GUT scale \cite{yanagida1,
HIY}.
The dynamical breaking of SU(5)$_{\rm GUT}$ also produces a pair of massless
composite color-triplet states simultaneously. A use of the missing partner
mechanism \cite{MNTY} yields a pair of massless Higgs doublets, giving large
masses to the color-triplet partners.

The model assumes, in addition to $Q$ and $\overline{Q}$, a hypercolor-singlet
chiral
multiplet $\Sigma$ in the adjoint representation of SU(5)$_{\rm GUT}$
to get rid of unwanted Nambu-Goldstone multiplets. The $\Sigma$ is easily
extended to the ({\bf 35}+{\bf1}) representation of the global SU(6)
by introducing a pair of {\bf 5} and {\bf 5}$^*$ and two singlets of
SU(5)$_{\rm GUT}$. Although this extended model is more complicated
than the previous one, it seems to suggest a more fundamental theory
at the short distance \cite{yanagida2}. Namely, this extended model may be
regarded
as the dual theory of an original SU(3) gauge theory with six flavors of quarks
$q$ and
$\overline{q}$, but without $\Sigma$ \cite{seiberg1}.

In this letter, however, we propose a different view on the extended model.
We show that a model with a specific choice of parameters in the
superpotential is identical to the low-energy effective theory of an \djn
SUSY model for the strongly interacting gauge sector.

The previous model \cite{yanagida1,HIY} is based on a supersymmetric
hypercolor SU(3)$_{\rm H}$ gauge theory with six flavors of quarks
$Q^A_\alpha$ in the {\bf 3} representation and antiquarks
$\overline{Q}_A^\alpha$ in the {\bf 3}$^*$ representation of the
hypercolor SU(3)$_{\rm H}$ ($\alpha$=1-3 and $A$=1-6). The first fives of
$Q^A_\alpha$ and $\overline{Q}_A^\alpha$ ($A$=1-5) transform as {\bf 5}$^*$
and {\bf 5} of the SU(5)$_{\rm GUT}$, respectively. In addition to the quarks
and antiquarks we introduce a hypercolor singlet $\Sigma^A_B$ ($A,B=1-5$) in
the {\bf 24}
representation of SU(5)$_{\rm GUT}$ to avoid unwanted Nambu-Goldstone
multiplets
\cite{yanagida1,HIY}.

It is a straightforward task to extend the $\Sigma^A_B$ to the {\bf 35}+{\bf 1}
representation of SU(6) under which $Q^A_\alpha$ and  $\overline{Q}_A^\alpha$
transform as {\bf 6}$^*$ and {\bf 6}, respectively. We introduce a
new pair of {\bf 5} + {\bf 5}$^*$ and two singlets of SU(5)$_{\rm GUT}$
and combine all of them
with the adjoint {\bf 24} to form {\bf 35} + {\bf 1} of the global SU(6).

We assume the following renormalizable superpotential;
\begin{eqnarray}
\label{potential1}
W &=& \lambda Q^A_\alpha \Sigma_A^B \overline{Q}_B^\alpha
    	+ \mu {\rm Tr} \Sigma
	+ m_1 {\rm Tr} \Sigma^2 + m_2 ({\rm Tr} \Sigma)^2
	+ \lambda_1 {\rm Tr} \Sigma^3 \nonumber\\
&&      + \lambda_2 {\rm Tr} \Sigma^2 {\rm Tr} \Sigma
        + \lambda_3 ({\rm Tr} \Sigma)^3~~~(A,B=1-6),
\end{eqnarray}
so that this model has a global SU(6)$\times$U(1) symmetry in the limit of the
SU(5)$_{\rm GUT}$ gauge coupling $g_5$ vanishing ($g_5=0$). The U(1) should
be gauged to have an unbroken U(1)$_{\rm Y}$ gauge symmetry
in the vacuum described bellow \cite{yanagida1,HIY}.

This model contains seven complex parameters ($\lambda$, $\mu$, $m_1$, $m_2$,
$\lambda_1$, $\lambda_2$, and $\lambda_3$) in the superpotential.
Two phases of $\lambda$ and $\mu^2$, for example, are absorbed to the fields
$Q^A_\alpha$ and $\Sigma^A_B$.
At the general point of the parameter space we always have the following
vacuum,
\begin{eqnarray}
\label{vacuum}
\overline{Q}_A^\alpha=\left(
\begin{array}{cccccc}
v&0&0&0&0&0\\
0&v&0&0&0&0\\
0&0&v&0&0&0
\end{array}\right),
&&
Q^A_\alpha =\left(
\begin{array}{ccc}
v&0&0\\
0&v&0\\
0&0&v\\
0&0&0\\
0&0&0\\
0&0&0
\end{array}
\right),
\nonumber \\
\Sigma^A_B = \left(
\begin{array}{cc}
\begin{array}{ccc}
0&&\\
&0&\\
&&0
\end{array}
&
0
\\
0
&
\begin{array}{ccc}
w&&\\
&w&\\
&&w
\end{array}
\end{array}
\right).
&&
\end{eqnarray}
In this vacuum our gauge group is broken down to the standard-model
gauge group as
\begin{eqnarray}
{\rm SU(5)_{GUT} \times U(1)_{\rm H} \times SU(3)_H }
&\rightarrow&
{\rm SU(3)_C \times SU(2)_L \times U(1)_Y},
\nonumber
\end{eqnarray}
where the unbroken gauge group SU(3)$_{\rm C}$ and U(1)$_{\rm Y}$
are linear combinations of the SU(3)$_{\rm H}$ and the SU(3) subgroup
of SU(5)$_{\rm GUT}$ and the U(1)$_{\rm H}$ and the U(1) subgroup
of SU(5)$_{\rm GUT}$, respectively. The GUT unification of the gauge coupling
constants in the standard model is practically achieved by taking sufficiently
large gauge coupling constants of SU(3)$_{\rm H}$$\times$U(1)$_{\rm H}$
\cite{yanagida1,HIY}.

In this vacuum (\ref{vacuum}) three pairs of the $Q^A_\alpha$ and
$\overline{Q}_A^\alpha$ ($A=4-6$) acquire masses $\lambda w$.
The integration of these massive quarks and antiquarks yields
an $N_f=N_C=3$ gauge theory where $N_f$ and $N_C$ are the numbers of
flavor and color, respectively. As shown by Seiberg \cite{seiberg2} the
classical vacuum
given in Eq.~(\ref{vacuum}) is slightly modified by instanton
effects in the case of $N_f=N_C$. However, the basic structure of the vacuum is
not changed
and the standard-model gauge group SU(3)$_{\rm C}\times$SU(2)$_{\rm
L}\times$U(1)$_{\rm Y}$
remains unbroken even in the quantum vacuum.

Since the global SU(6) (in the limit $g_5$=0) is broken down to SU(3)$_{\rm
C}\times$SU(3)
in this vacuum, we have three true and two pseudo Nambu-Goldstone multiplets.
The two true Nambu-Goldstone multiplets transforming as ({\bf 3},{\bf 2})
and ({\bf 3}$^*$,{\bf 2}) under SU(3)$_{\rm C}\times$SU(2)$_{\rm L}$
are absorbed to the SU(5)$_{\rm GUT}$ gauge multiplets and the last one to the
broken U(1) gauge multiplet.
Thus, there remains a pair of pseudo Nambu-Goldstone multiplets which
transforms as
{\bf 3} and {\bf 3}$^*$ of SU(3)$_{\rm C}$. As shown in the previous
paper \cite{yanagida1,HIY}, these {\bf 3} and {\bf 3}$^*$ acquire GUT-scale
masses together with the {\bf 3} and {\bf 3}$^*$ in the standard
Higgs multiplets $H_A$({\bf 5}) and $\overline{H}^A$({\bf 5}$^*$) of
SU(5)$_{\rm GUT}$. Thus, there is a pair of Higgs doublets left in the
massless spectrum. It is easily proved by using the SUSY nonrenormalization
theorem \cite{SII} that this pair of Higgs doublets is exactly massless even
in the quantum vacuum as far as the SUSY is unbroken.

Now, we show our main point in this letter. There is a peculiar
hypersurface in the parameter space of the extended model, on
which U(1)$_{\rm R}$ symmetry is restored. This surface is defined by
$\mu=\lambda_1=\lambda_2=\lambda_3=0$ and hence it consists of
only 3 parameters $\lambda$, $m_1$, and $m_2$ in Eq.~(\ref{potential1}).
Under the U(1)$_{\rm R}$, each fields, $Q^A_\alpha$, $\overline{Q}_A^\alpha$,
and $\Sigma^A_B$ transform as
\begin{eqnarray}
Q^A_\alpha (\theta)
&\rightarrow&
{\rm e}^{-\frac{i}2\alpha} Q^A_\alpha (\theta{\rm e}^{i\alpha}),
\nonumber\\
\overline{Q}_A^\alpha(\theta)
&\rightarrow&
{\rm e}^{-\frac{i}2\alpha} \overline{Q}_A^\alpha (\theta{\rm e}^{i\alpha}),
\nonumber\\
\Sigma^A_B(\theta)
&\rightarrow&
{\rm e}^{-i\alpha} \Sigma^A_B(\theta{\rm e}^{i\alpha}).
\end{eqnarray}

On this hypersurface we have the vacuum defined in Eq.~(\ref{vacuum})
only when $m_2=-\frac13 m_1$ is satisfied. Thus a fine tuning of the
parameters is required to obtain the desired vacuum given in
Eq.~(\ref{vacuum}). However, we show that such a specific point in the
parameter space
corresponds to a low-energy effective theory of an \djn SUSY gauge model
for the strongly interacting sector.

The \djn SUSY model discussed by Leigh and Strassler \cite{LS} contains an
N=2 gauge multiplet of SU(3)$_{\rm H}$
and six (=2$N_C$) hypermultiplets ($Q^A_\alpha, \overline{Q}_A^\alpha$). The
N=2 gauge multiplet consists of an SU(3)$_{\rm H}$ adjoint chiral
superfield $X_\beta^\alpha$ and the SU(3)$_{\rm H}$ gauge multiplet in the
N=1 SUSY theory. Then, the superpotential is written in terms of the N=1
SUSY fields as
\begin{equation}
\label{n=2potential}
W=g Q^A_\alpha X_\beta^\alpha \overline{Q}_A^\beta,
\end{equation}
where $g$ is the SU(3)$_{\rm H}$ gauge coupling constant. The \djn SUSY
model is defined by introducing the mass term for  $X_\beta^\alpha$
\cite{yanagida2},
\begin{equation}
\label{n=2potential1}
\delta W= \frac{m_X}2 {\rm Tr} X^2,
\end{equation}
which breaks the N=2 SUSY down to the N=1.

The integration of $X_\beta^\alpha$ leads to an effective superpotential
\begin{eqnarray}
W^\prime &=& -\frac{g^2}{m_X}
\left(Q^A_\alpha \overline{Q}_A^\beta\right)
{\rm T}^{i\alpha}_{\beta}{\rm T}^{i\gamma}_{\delta}
\left(Q^B_\gamma \overline{Q}_B^\delta\right)
\nonumber\\
&=&
\frac{g^2}{2 m_X}
\left(
\left(Q^A_\alpha \overline{Q}_B^\alpha\right)
\left(Q^B_\beta \overline{Q}_A^\beta\right)
-\frac13
\left(Q^A_\alpha \overline{Q}_A^\alpha\right)
\left(Q^B_\beta \overline{Q}_B^\beta\right)
\right),
\end{eqnarray}
where T$^i$ is a generator matrix of the SU(3)$_{\rm H}$ Lie algebra.
This superpotential
is rewritten by using an auxiliary field $\Sigma^A_B$ as
\begin{eqnarray}
W^\prime &=& \lambda Q^A_\alpha \Sigma^B_A \overline{Q}_B^\alpha
+ \frac{\lambda^2 m_X}{2 g^2}
\left[ {\rm Tr}\Sigma^2 - \frac13 ({\rm Tr}\Sigma)^2 \right].
\end{eqnarray}
This corresponds exactly to the superpotential at the specific point in the
parameter space of our extended model on which the desired vacuum given in
Eq.~(\ref{vacuum}) is obtained. In fact, we drive F-term flatness
conditions for vacua directly from the
superpotential (\ref{n=2potential}) and (\ref{n=2potential1}) as,
\begin{eqnarray}
X_\alpha^\beta \overline{Q}_A^\alpha &=& Q^A_\alpha X_\beta^\alpha = 0,
\\
m_X X_\alpha^\beta &=& -g (
\overline{Q}_A^\beta Q^A_\alpha
-\frac13 \delta_\alpha^\beta {\rm Tr}\overline{Q}Q),
\end{eqnarray}
which yield the solution for $Q^A_\alpha$ and $\overline{Q}_A^\alpha$
given in Eq.~(\ref{vacuum}) together with D-term flatness conditions.

At the short distance the $\Sigma^A_B$ is an auxiliary field of
dimension two in the \djn SUSY model while the $\Sigma^A_B$ in the extended
model is a propagating canonical field of dimension one. However, it is very
much plausible that the radiative
corrections generate the kinetic term for the auxiliary field $\Sigma^A_B$
at the considerably long distance like at the GUT scale. Therefore, we believe
that both theories become indistinguishable  at the low energies as point out
in Ref.~\cite{SII}\footnote{
Precisely speaking, however, dynamics in our model is not the same as that
in the \djn SUSY model in Ref.~\cite{SII} because of the presence of
U(1)$_{\rm H}$ gauge interaction, even when one turns off the weakly
interacting SU(5)$_{\rm GUT}$. Nevertheless, it is amusing to note that the
gauge coupling constant $g_{\rm 1H}$ of the U(1)$_{\rm H}$ is vanishing at
the very long distance and hence the argument in Ref.~\cite{SII} may be
applicable even in our \djn SUSY model if one could take such a long distance.
}

Since the \djn
SUSY model does not need a fine-tuning of parameter to have the desired
vacuum, we consider that the extended model at a specific point in the
parameter space is a low-energy effective theory of the more fundamental
\djn SUSY model.

We should note that there is a massless Nambu-Goldstone
multiplet corresponding to the U(1)$_{\rm R}$ breaking. As a consequence
there is a exactly flat direction keeping the vacuum-expectation value $v$
in Eq.~(\ref{vacuum}) undetermined unless the explicit breaking term of the
U(1)$_{\rm R}$ is introduced.

Finally we comment on the nucleon's instability. Unlike the previous model
\cite{yanagida1,HIY} we do not have a suppression mechanism for the dangerous
d=5 operators of the nucleon decay in the present model. Therefore, we hope
that the coming nucleon-decay experiments will select one of these two
different models.

{}~~\\
\underline{\it Acknowledgment}\\
One of the authors (T.Y.) thanks T. Hotta and K.I. Izawa for useful discussions
on the extension of the original model in Ref.~\cite{yanagida1,HIY} and
S. Iso for informing the presence of Ref.~\cite{LS}. We also thank the
high-energy theory group at ITP, Santa Barbara  California, for their
hospitality. It was during the period of stay at ITP that the ideas discussed
in this paper germinated. This work was supported by National Foundation
under grant no. PHY94-07194.

\newpage
%
%
\newcommand{\Journal}[4]{{\sl #1} {\bf #2} {(#3)} {#4}}
\newcommand{\APJ}{Ap. J.}
\newcommand{\CJP}{Can. J. Phys.}
\newcommand{\NC}{Nuovo Cimento}
\newcommand{\NP}{Nucl. Phys.}
\newcommand{\PL}{Phys. Lett.}
\newcommand{\PR}{Phys. Rev.}
\newcommand{\PRep}{Phys. Rep.}
\newcommand{\PRL}{Phys. Rev. Lett.}
\newcommand{\PTP}{Prog. Theor. Phys.}
\newcommand{\SJNP}{Sov. J. Nucl. Phys.}
\newcommand{\ZP}{Z. Phys.}

\end{document}